\documentclass{svmult}

\usepackage{mathptmx}
\usepackage{helvet}
\usepackage{courier}
\usepackage{makeidx}
\usepackage{graphicx}
\usepackage{multicol}
\usepackage[bottom]{footmisc}
\usepackage{hyperref}
\usepackage{rotating}

\makeindex

\begin{document}

\title*{Mapping the Curricular Structure and Contents of Network Science Courses}
\author{Hiroki Sayama}
\institute{Hiroki Sayama \at Center for Collective Dynamics of Complex
  Systems, Binghamton University, State University of New York,
  Binghamton, NY 13902-6000, USA / Center for Complex Network
  Research, Northeastern University, Boston, MA 02115, USA / Faculty
  of Commerce, Waseda University, Shinjuku, Tokyo 169-8050, Japan.
  \email{sayama@binghamton.edu}}

\maketitle

\abstract*{As network science has matured as an established field of
  research, there are already a number of courses on this topic
  developed and offered at various higher education institutions,
  often at postgraduate levels. In those courses, instructors 
  adopted different approaches with different focus areas and
  curricular designs. We collected information about 30 existing
  network science courses from various online sources, and analyzed
  the contents of their syllabi or course schedules. The topics and
  their curricular sequences were extracted from the course
  syllabi/schedules and represented as a directed weighted graph,
  which we call the {\em topic network.} Community detection in the
  topic network revealed seven topic clusters, which matched
  reasonably with the concept list previously generated by students
  and educators through the Network Literacy initiative. The minimum
  spanning tree of the topic network revealed typical flows of
  curricular contents, starting with examples of networks, moving onto
  random networks and small-world networks, then branching off to
  various subtopics from there. These results illustrate the current
  state of consensus formation (including variations and
  disagreements) among the network science community on what should be
  taught about networks and how, which may also be informative for
  K--12 education and informal education.}

\abstract{As network science has matured as an established field of
  research, there are already a number of courses on this topic
  developed and offered at various higher education institutions,
  often at postgraduate levels. In those courses, instructors 
  adopted different approaches with different focus areas and
  curricular designs. We collected information about 30 existing
  network science courses from various online sources, and analyzed
  the contents of their syllabi or course schedules. The topics and
  their curricular sequences were extracted from the course
  syllabi/schedules and represented as a directed weighted graph,
  which we call the {\em topic network.} Community detection in the
  topic network revealed seven topic clusters, which matched
  reasonably with the concept list previously generated by students
  and educators through the Network Literacy initiative. The minimum
  spanning tree of the topic network revealed typical flows of
  curricular contents, starting with examples of networks, moving onto
  random networks and small-world networks, then branching off to
  various subtopics from there. These results illustrate the current
  state of consensus formation (including variations and
  disagreements) among the network science community on what should be
  taught about networks and how, which may also be informative for
  K--12 education and informal education.}

\section{Introduction}

Network science has grown at a rapid pace over the last few decades,
producing several major international conferences, scientific
journals, research communities, and even academic degree programs
\cite{barabasi2016network}. As it has matured as an established field
of research, there are already a number of courses on this topic
developed and offered at various higher education institutions, often
at postgraduate levels. Those courses are delivered in several
different departments/disciplines with their respective emphases, such
as mathematics, computer science, physics, sociology, political
science, management science, systems science, biology, medicine, and
in other more interdisciplinary settings as well.

In those recently developed network science courses, instructors 
adopted different approaches with different focus areas and curricular
designs, depending on their backgrounds, knowledge, and objectives. It
should be of particular interest to the network science community to
investigate what are agreed or disagreed upon among those instructors
on the choices of topics and the curricular flows that go through
those topics in a sequential instruction. To the best of our
knowledge, there is no prior literature on such a systematic analysis
of network science course contents.

The study presented in this chapter aims to collect and organize the
information about a number of existing network science courses,
generate ``maps'' of their curricular structures, and identify a set
of commonly used curricular contents and typical flows of instruction
that connect those contents. Information about course contents were
extracted from the online syllabi or schedules of the network science
courses and were modeled as a directed weighted graph, to which
several network analysis methods were applied to reveal underlying
curricular structure. Potential directions of further improvement of
network science curriculum design are also discussed based on the
results.

\section{Data Collection}

We gathered information about existing network science courses from
online sources, using the following two websites as the main
starting points:
\begin{itemize}
\item Complexity Explorer\\\url{https://www.complexityexplorer.org/}
\item Awesome Network Analysis\\\url{https://github.com/briatte/awesome-network-analysis}
\end{itemize}
From these websites we collected the syllabi or course schedules of
several dozens of English-based courses that included topics related
to networks. As our objective was to analyze the curricular structure
of ``network science'' as an interdisciplinary field of research, we
excluded the following types of courses from our analysis:
\begin{enumerate}
\item Purely mathematical graph theory courses
\item Statistics courses that included network analysis only briefly
\item Courses on narrowly defined applications (e.g., political
  analysis, genomic analysis)
\item Special topics/seminar courses
\end{enumerate}
As a result, we selected the 30 courses shown in Table
\ref{tab:courses} as the data sources for our study.

\begin{table}
\centering
\caption{URLs of 30 courses from which curricular information was
  corrected for this study. The original URLs used for data collection
  in April--May 2016 are shown here, some of which may have been
  updated since then or may no longer be available. Note that some
  institutions are represented multiple times in this list, while
  others appear only once. This may have an biasing effect on the
  results of analysis.}
\noindent\rule{\columnwidth}{.3pt}
\begin{enumerate}
\item \url{http://barabasi.com/book/network-science}
\item \url{http://bingweb.binghamton.edu/~sayama/SSIE641/}
\item \url{http://faculty.nps.edu/rgera/MA4404.html}
\item \url{http://hornacek.coa.edu/dave/Teaching/Networks.11/}
\item \url{http://mae.engr.ucdavis.edu/dsouza/mae298}
\item \url{http://networksatharvard.com/}
\item \url{http://ocw.mit.edu/courses/economics/14-15j-networks-fall-2009/}
\item \url{http://ocw.mit.edu/courses/media-arts-and-sciences/mas-961-networks-complexity-and-its-applications-spring-2011/}
\item \url{http://perso.ens-lyon.fr/marton.karsai/Marton_Karsai/complexnet.html}
\item \url{https://cns.ceu.edu/node/31544}
\item \url{https://cns.ceu.edu/node/31545}
\item \url{https://cns.ceu.edu/node/38501}
\item \url{https://courses.cit.cornell.edu/info2040_2015fa/}
\item \url{https://iu.instructure.com/courses/1491418/assignments/syllabus}
\item \url{https://sites.google.com/a/yale.edu/462-562-graphs-and-networks/}
\item \url{https://www0.maths.ox.ac.uk/courses/course/28833/synopsis}
\item \url{https://www.coursera.org/course/sna}
\item \url{https://www.sg.ethz.ch/media/medialibrary/2014/11/syllabus-cn15.pdf}
\item \url{http://tuvalu.santafe.edu/~aaronc/courses/5352/}
\item \url{http://web.stanford.edu/class/cs224w/handouts.html}
\item \url{http://web.stanford.edu/~jugander/mse334/}
\item \url{http://www2.warwick.ac.uk/fac/cross_fac/complexity/study/msc_and_phd/co901/}
\item \url{http://www.ait-budapest.com/structure-and-dynamics-of-complex-networks}
\item \url{http://www.cabdyn.ox.ac.uk/Network%20Courses/SNA_Handbook%202013-14.pdf}
\item \url{http://www.cc.gatech.edu/~dovrolis/Courses/NetSci/}
\item \url{http://www.columbia.edu/itc/sociology/watts/w3233/}
\item \url{http://www.cse.unr.edu/~mgunes/cs765/}
\item \url{http://www-personal.umich.edu/~mejn/courses/2015/cscs535/index.html}
\item \url{http://www.stanford.edu/~jacksonm/291syllabus.pdf}
\item \url{http://www.uvm.edu/~pdodds/teaching/courses/2016-01UVM-303/}
\end{enumerate}
\noindent\rule{\columnwidth}{.3pt}
\label{tab:courses}
\end{table}

Data collection was conducted manually by the author in April--May
2016. Network science-related topics were extracted from each of the
data sources and were grouped by instructional modules shown in the
syllabi/schedule. All of the extracted topics were converted to
lowercase letters without diacritics to facilitate text
processing. The topics were also often
normalized/edited/reworded/aggregated at the discretion of the author,
to make the vocabulary consistent throughout the analysis. The cleaned
final data set (including the rewording rules used in this study) is
available from figshare \cite{figshare}.

\section{Methods of Analysis}

\begin{figure}[tbp]
\centering
\includegraphics[width=\columnwidth]{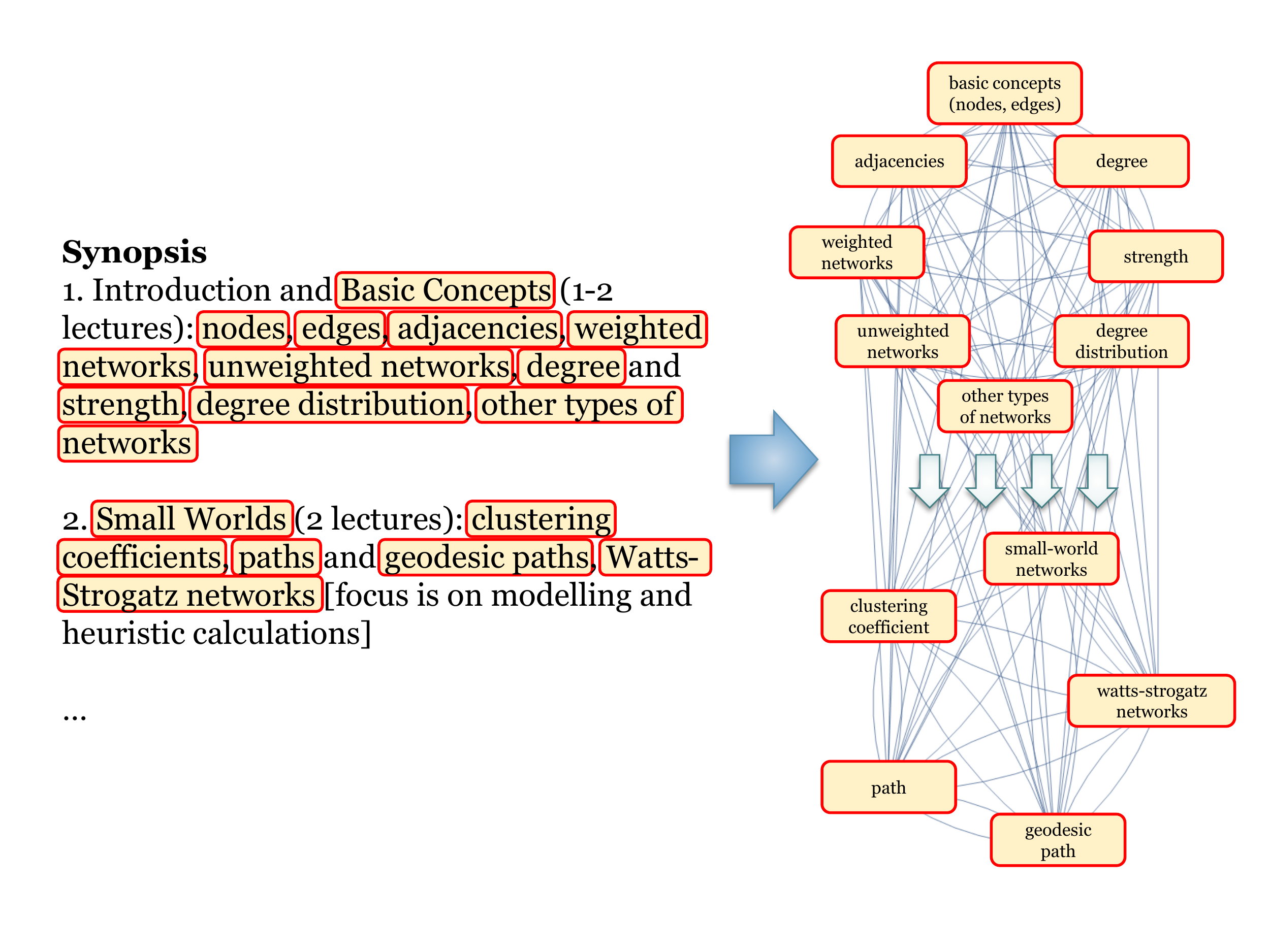}
\caption{Schematic illustration showing how the edges in the topic
  network were created from course syllabi/schedules. Left: An excerpt
  from a sample network science course syllabus (from Mason Porter's
  course
  \url{https://www0.maths.ox.ac.uk/courses/course/28833/synopsis};
  also see \cite{porter2017course}), in which extracted topics are
  highlighted. Right: A subgraph of the topic network created from the
  excerpt on the left. Topics that appear in the same curricular
  module were connected to each other with bidirectional
  edges. Directed edges were also created from topics covered in the
  previous module (top) to those covered in the subsequent module
  (bottom) to represent curricular flows. The extracted topics were
  often normalized/edited/reworded/aggregated at the discretion of the
  author, to make the vocabulary consistent throughout the analysis.}
\label{fig:method}
\end{figure}

The topics and their curricular sequences extracted from the course
syllabi/schedules were initially represented as a directed multigraph
by the following procedure (also see Fig.~\ref{fig:method}):
\begin{enumerate}
\item Connect topics that appear in the same curricular module to each
  other with bidirectional edges, to form a fully connected cluster of
  topics.
\item Connect topics covered in the previous module to those covered
  in the subsequent module with directed edges, to represent
  curricular flows.
\end{enumerate}
These steps were repeated for all curricular modules in all of the
courses. After this edge construction process was over, multiple edges
that shared the same origin-destination pair were replaced by a single
directed weighted edge with the multiplicity of the original edges as
the weight. The result was obtained as a single large directed
weighted graph, which we call the {\em topic network} hereafter. This
topic network was analyzed using several different methods.

First, the distribution of instructional attention/emphasis in the
current network science courses was characterized by measuring the
absolute frequencies of appearance of topics in the original data
set. We did not use degree or other centrality measures in the topic
network for this purpose, because, according to the procedure of
network construction used in this study (Fig.~\ref{fig:method}), each
topic's in- and out-degrees are greatly influenced by the numbers of
other topics in previous and next curricular modules, respectively.

Next, all of the edges whose weight was two or below were removed from
the topic network and only the largest strongly connected component
was kept for the rest of the analysis, in order to improve the
robustness of the findings by focusing on the essential main body of
the topic network. Communities of topics were detected by applying the
modularity maximization method
\cite{blondel2008fast,fortunato2016community} to the topic
network. Finally, the edge weights were inverted from the original
ones so they would represent distance (not strength) of connections,
and then the minimum spanning tree (i.e., a tree that reaches all of the
nodes with the minimal sum of edge weights) \cite{graham1985history}
of this weight-inverted topic network was computed to reveal typical
flows of instruction going through various network science topics.

For all of these analyses and visualizations, we used Wolfram Research
Mathematica 11.1.1.

\section{Results}

\begin{figure}[tbp]
\centering
\includegraphics[width=\columnwidth]{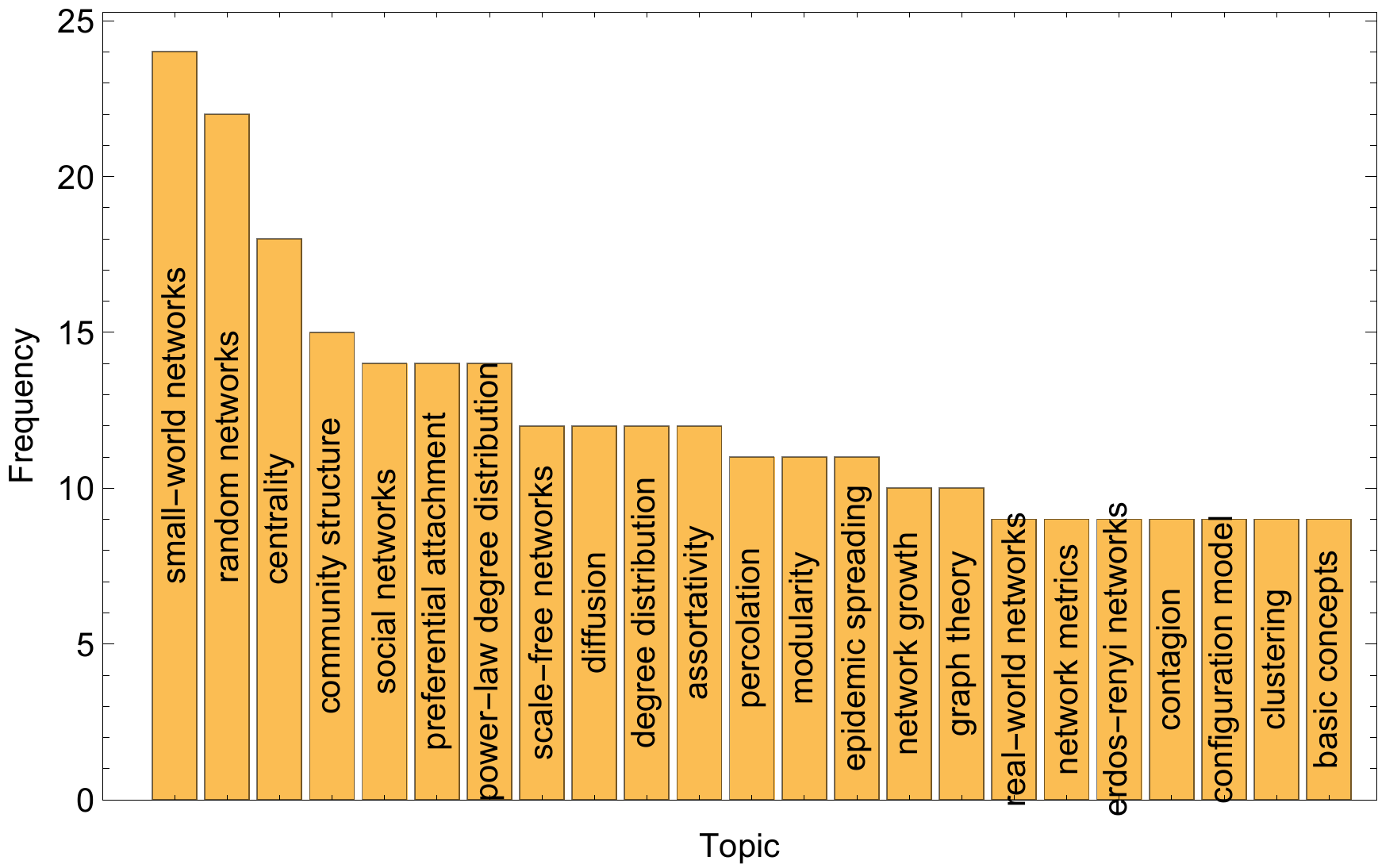}
\caption{Frequencies of the top 20 topics that appeared most
  frequently in the 30 course syllabi/schedules (ties were included so
  a total of 23 topics appear in this chart). Note that all of the
  extracted topics were converted to lowercase letters without
  diacritics to facilitate text processing (this applies to the
  other figures as well).}
\label{fig:top-topics}
\end{figure}

\begin{figure}[tbp]
\centering
\includegraphics[width=0.8\columnwidth]{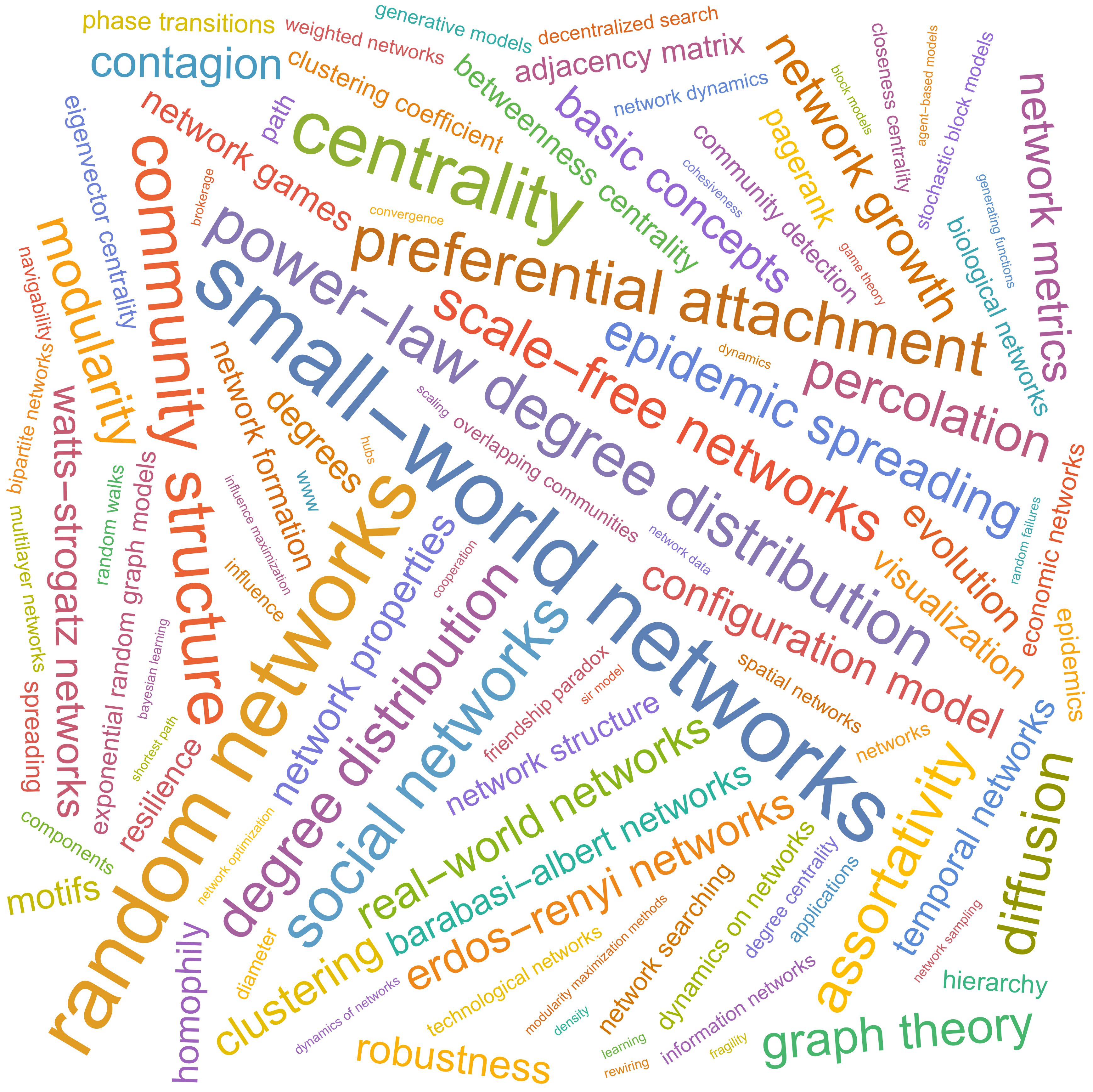}
\caption{Visualization of topic frequencies in the 30 course
  syllabi/schedules as a word cloud. Font sizes are set proportional
  to the square roots of topic frequencies.}
\label{fig:word-cloud}
\end{figure}

Figure \ref{fig:top-topics} shows the top 20 topics that appeared most
frequently in the collected course syllabi/schedules. The topic
``small-world networks'' appeared most frequently in our analysis,
probably because this topic was covered widely in various disciplines,
including math/physics/computer science, social/economic/political
sciences, psychology/neuroscience, and some others. ``Random
networks'', ``centrality'', and other well-known topics are also
represented in this list. A larger set of topics is visualized as
a word cloud in Fig.~\ref{fig:word-cloud}.

\begin{figure}[tbp]
\centering
\includegraphics[width=\columnwidth]{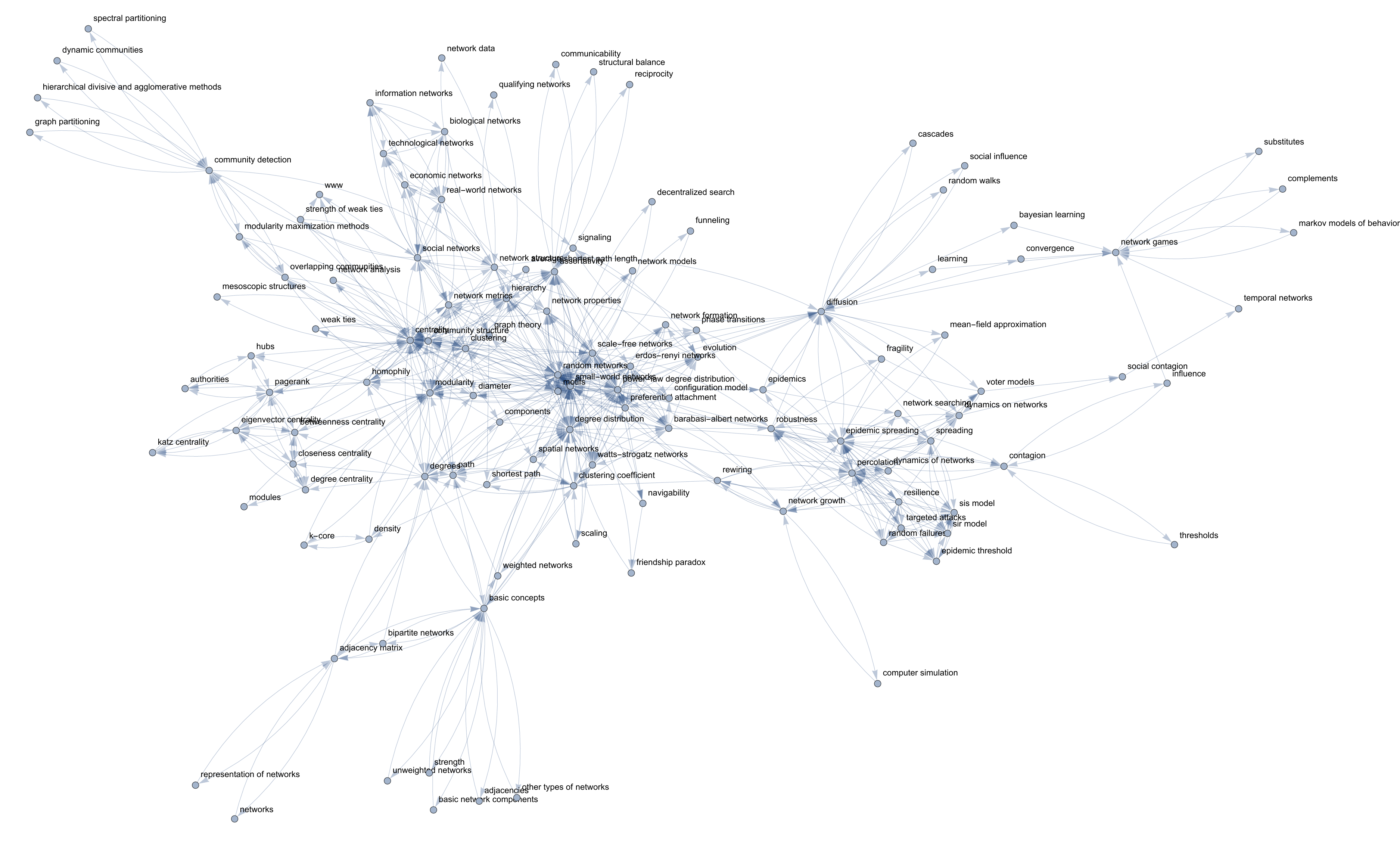}
\caption{Visualization of the topic network after edge weight
  thresholding. Only the largest strongly connected component is
  shown. Edge weights are ignored to simplify the visualization.}
\label{fig:topic-network2}
\end{figure}

Figure \ref{fig:topic-network2} shows a visualization of the filtered
topic network after edge weight thresholding and extraction of the
largest strongly connected component. High-resolution versions of this
and other visualizations are available from figshare
\cite{figshare}. While the original topic network included 505 topics,
the filtered one included 121. The latter was more focused on
essential, frequently covered topics than the original, and thus we
used the filtered one for the rest of the analysis.


\begin{figure}[tbp]
\centering
\begin{turn}{90}
\includegraphics[height=\columnwidth]{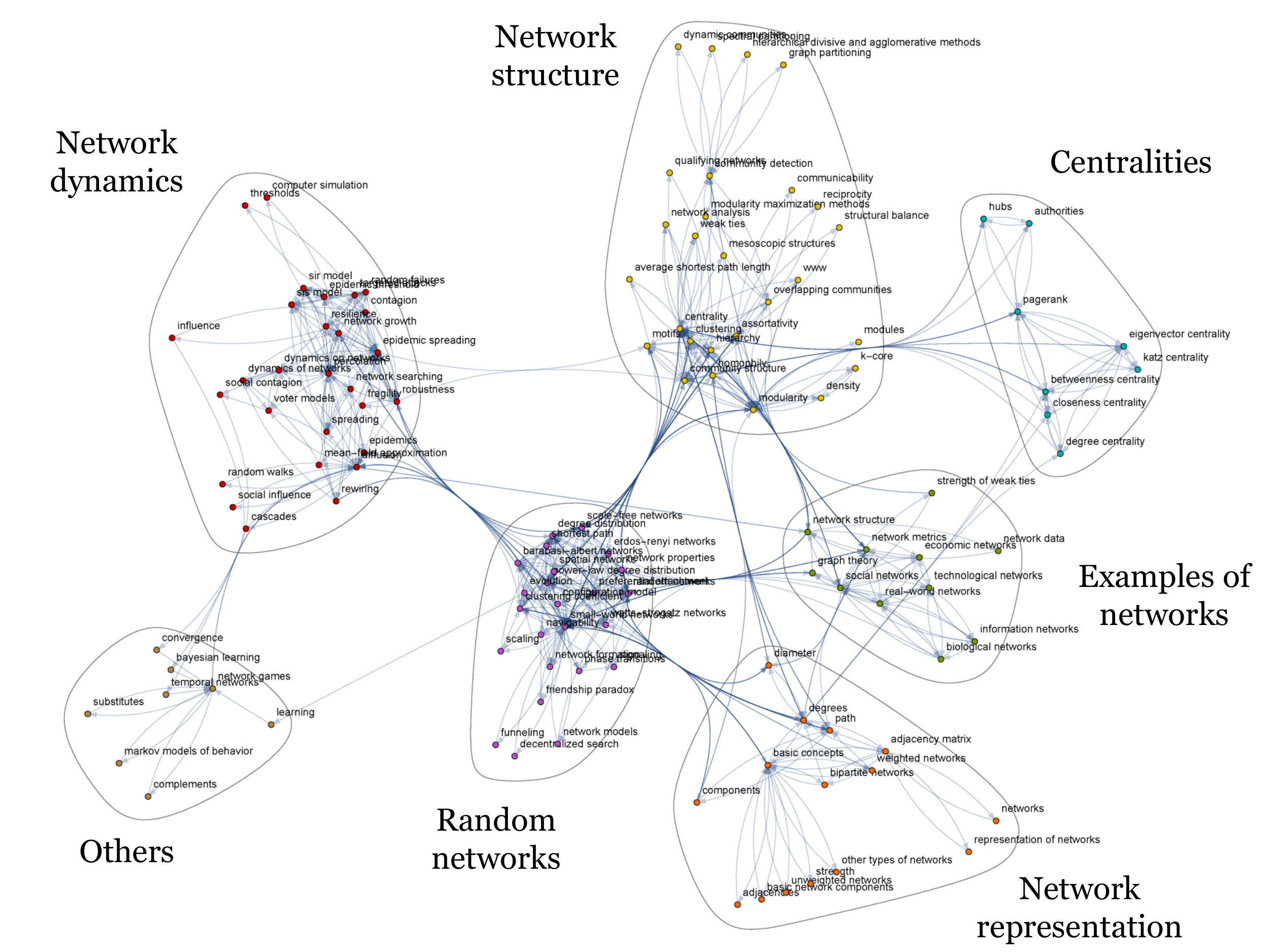}
\end{turn}
\caption{(Rotated) Communities detected by applying the modularity
  maximization method to the topic network. Seven topic clusters 
  were detected: (1) {\em examples of networks}, (2) {\em network
    representation}, (3) {\em random networks}, (4) {\em network
    structure}, (5) {\em centralities}, (6) {\em network dynamics}, and
  (7) {\em others}.}
\label{fig:communities}
\end{figure}

Figure \ref{fig:communities} shows the communities of topics detected
by applying the modularity maximization method to the filtered topic
network. Seven topic clusters were detected. Although characterizing
each cluster with an appropriate label was a challenging task, we
reviewed the content of each cluster and came up with the following
characterization of the seven clusters:
\begin{enumerate}
\item {\em Examples of networks (middle-right).} This cluster includes
  concrete examples of networks, such as social networks, economic
  networks, biological networks, technological networks, and
  information networks.
\item {\em Network representation (bottom-right).} This cluster
  includes fundamental concepts and terminologies about representation
  of networks, such as basic network components, adjacencies, path,
  degree, strength, etc.
\item {\em Random networks (bottom-center).} This cluster is the most
  dense and the most difficult to label. It includes a wide variety of
  topics, and many of them had strong connections to other
  communities. However, it uniquely includes several major random
  network models (e.g., Erd\H{o}s--R\'enyi networks, small-world
  networks, Barab\'asi--Albert networks, preferential attachment,
  etc.). Therefore we tentatively call this cluster ``random
  networks''. It is clearly the core part of this topic community map.
\item {\em Network structure (top-center).} This cluster includes
  concepts about network structure and tools to analyze it, such as
  clustering, path length, modularity, community detection, k-core,
  etc.
\item {\em Centralities (top-right).} This relatively small cluster
  has a clear focus on centrality measures.
\item {\em Network dynamics (top-left).} This cluster includes various
  dynamical concepts that are typically discussed in dynamical
  systems, stochastic/probabilistic systems, and statistical physics,
  such as spreading/contagion, influence, and dynamics on/of networks.
\item {\em Others (bottom-left).} This small cluster includes
  miscellaneous topics that do not appear to have a common theme
  (e.g., learning, network games, temporal networks).
\end{enumerate}

The cluster of random networks occupies a central position in this
map, to which most other clusters are attached with varying degrees of
connection strength. The connections are particularly strong between
random networks and network structure, as well as between random
networks and network dynamics, indicating their strong linkages in the
core curricula of network science courses.

We compared the topic clusters identified above with the essential
concepts generated by students and educators through the Network
Literacy initiative \cite{networkliteracy,sayama2016essential} (Table
\ref{tab:matching}). They matched reasonably regarding examples of
networks, network representation, network structure/centralities, and
network dynamics. In the meantime, the cluster of random networks does
not have a counterpart in the essential concepts list, probably
because the topics covered in this cluster are somewhat at advanced
levels and may not be suitable for secondary education or general
public. On the other hand, the essential concepts about visualization
and computer technology (4 and 5 in the second column of Table
\ref{tab:matching}) were not well represented in the topic communities
seen in Fig.~\ref{fig:communities}. This finding coincides with the
fact that those two essential concepts were suggested and emphasized
by NetSci High \cite{cramer2015netsci} students, not by network
science researchers, when the Network Literacy booklet was developed
\cite{sayama2016essential}. This may indicate that the current
curricular structure of network science courses are likely not
spending sufficient time or resource to cover computational tools and
visualization methods, even though they could be essential for
students' learning of networks. A potential factor contributing to
this gap may be that many of the courses analyzed here are at advanced
graduate levels, where computational methods and visualization tools
may not be part of the core curricular content.

\begin{table}[tbp]
\caption{Comparison between the topic clusters revealed in
  Fig.~\ref{fig:communities} and the essential concepts developed in
  the Network Literacy initiative
  \cite{networkliteracy,sayama2016essential}.}
\centering
\begin{tabular}{p{1.3in}p{2.7in}c}
\hline
Topic cluster detected & Essential concept given in Network Literacy & Matched? \\
\hline
1. Examples of networks & 1. Networks are everywhere.  & Yes \\
2. Network representation & 2. Networks describe how things connect and interact. & Yes \\
3. Random networks & {\em (missing)} & {\bf No} \\
4. Network structure & 3. Networks can help reveal patterns. & Yes \\
5. Centralities &  {\em (covered in 3)} & Yes \\
{\em (missing)} & 4. Visualizations can help provide an understanding of networks. & {\bf No} \\
{\em (missing)} & 5. Today's computer technology allows you to study real-world networks. & {\bf No} \\
{\em (covered in 1?)} & 6. Networks help you to compare a wide variety of systems. & Yes? \\
6. Network dynamics & 7. The structure of a network can influence its state and vice versa. & Yes \\
7. Others & {\em (covered in 7?)} & Yes? \\
\hline
\end{tabular}
\label{tab:matching}
\end{table}

\begin{figure}
\centering
\begin{turn}{90}
\includegraphics[height=\columnwidth]{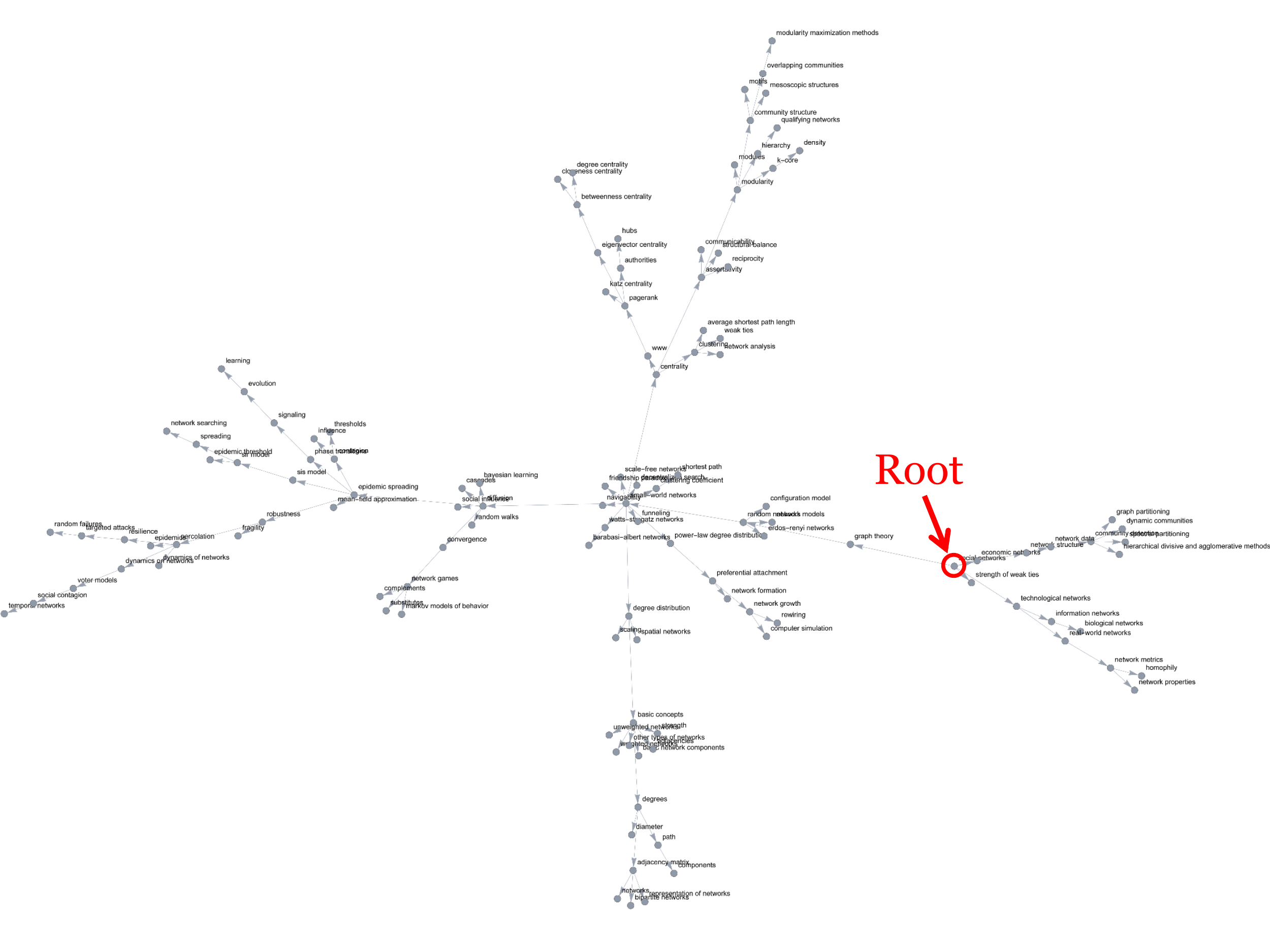}
\end{turn}
\caption{(Rotated) Minimum spanning tree of the topic network with
  inverted edge weights. The root of the tree is indicated by a red
  circle. This spanning tree presents a sample ``backbone'' of
  curricular flows among various network science
  concepts. High-resolution version is available from figshare \cite{figshare}.}
\label{fig:whole-map}
\end{figure}

Finally, Figure \ref{fig:whole-map} presents the minimum spanning tree
of the topic network with inverted edge weights. This map shows the
curricular structure of network science courses in greater detail with
sequential relationships, revealing a possible ``backbone'' of
curricular flows among various network science concepts. The root of
the tree is located near the right side of the map, starting with
social networks. From there, several curricular flows can be
identified on this map. Details are explained below with enlarged
portions of the map, which turn out to bear a good correspondence with
the topic clusters detected in Fig.~\ref{fig:communities}.

\begin{figure}[tbp]
\centering
\includegraphics[width=\columnwidth]{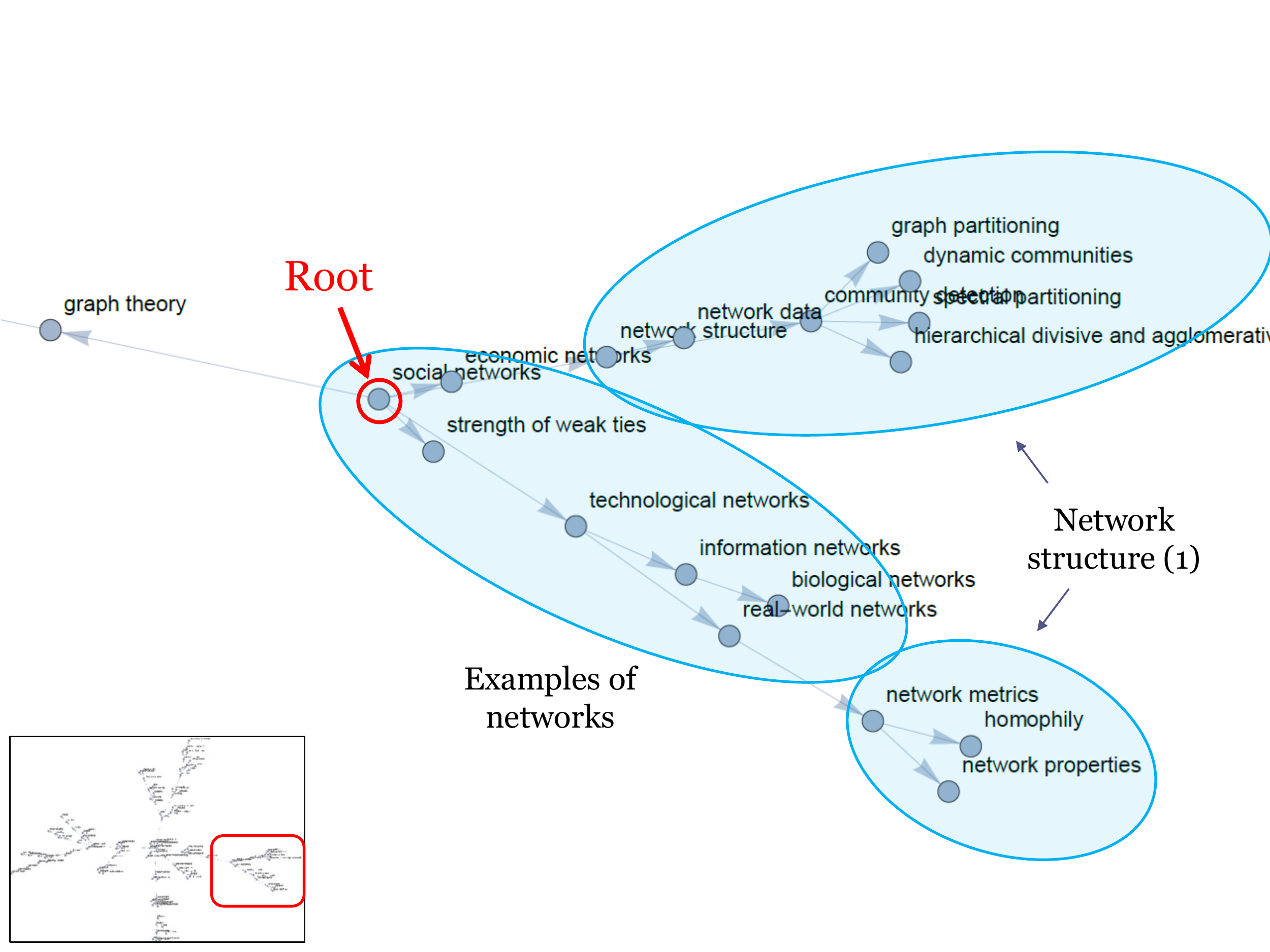}
\caption{Enlarged right portion of the spanning tree shown in
  Fig.~\ref{fig:whole-map}. Two branches, covering {\em examples of
    networks} and {\em network structure}, extend from the root of the
  spanning tree.}
\label{fig:map1}
\end{figure}

Figure \ref{fig:map1} shows the right portion of
Fig.~\ref{fig:whole-map}, in which the root of the spanning tree,
social networks, is located in the middle. Two branches are shown in
this figure, in addition to another path going from the root
leftward. The first branch (lower one) includes topics such as
technological networks, information networks, biological networks, and
real-world networks, which clearly correspond to the topic cluster of
{\em examples of networks}. The other branch (upper one) includes
network data, community detection, partitioning, and other related
topics, which could be summarized as {\em network structure}, together
with a few other topics that show up at the tip of the first
branch. This area of the tree appears to be an introductory part of
the curricular structure.

\begin{figure}[tbp]
\centering
\includegraphics[width=\columnwidth]{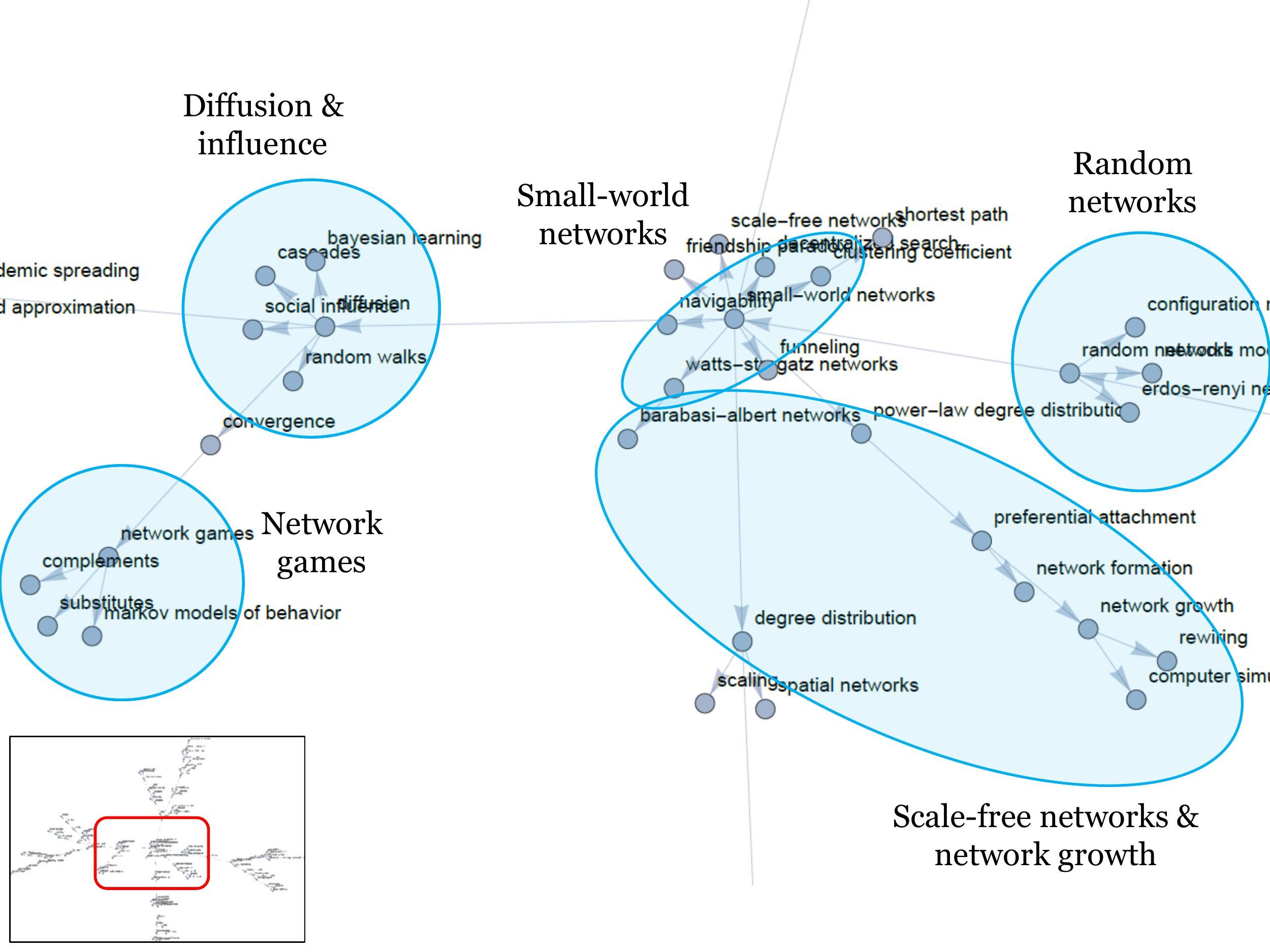}
\caption{Enlarged central portion of the spanning tree shown in
  Fig.~\ref{fig:whole-map}. The curricular flow originating at the
  root (not shown in this figure) comes from the right, goes through
  {\em random networks}, then reaches {\em small-world networks}. From
  there several outgoing branches emanate, including {\em scale-free
    networks \& network growth} and {\em diffusion \& influence}; the
  latter is followed by {\em network games.}}
\label{fig:map2}
\end{figure}

Figure \ref{fig:map2} shows the central portion of
Fig.~\ref{fig:whole-map}, which is the busiest area in the spanning
tree where a number of new concepts and models are introduced. The
curricular flow that originated in the root comes from the right, and
first goes through {\em random networks}, where purely random network
models such as Erd\H{o}s--R\'enyi models and configuration models are
introduced. Then it reaches {\em small-world networks} that serves as
the crux of the whole spanning tree. The observed importance of the
small-world networks in the curricular flow agrees with its highest
frequency seen in Fig.~\ref{fig:top-topics}. From there, several
different flows branch off toward various subtopics, most notably {\em
  scale-free networks \& network growth} that goes down to the
right. Other topics shown in this figure are {\em diffusion \&
  influence} and {\em network games} to the left.

\begin{figure}[tbp]
\centering
\includegraphics[width=\columnwidth]{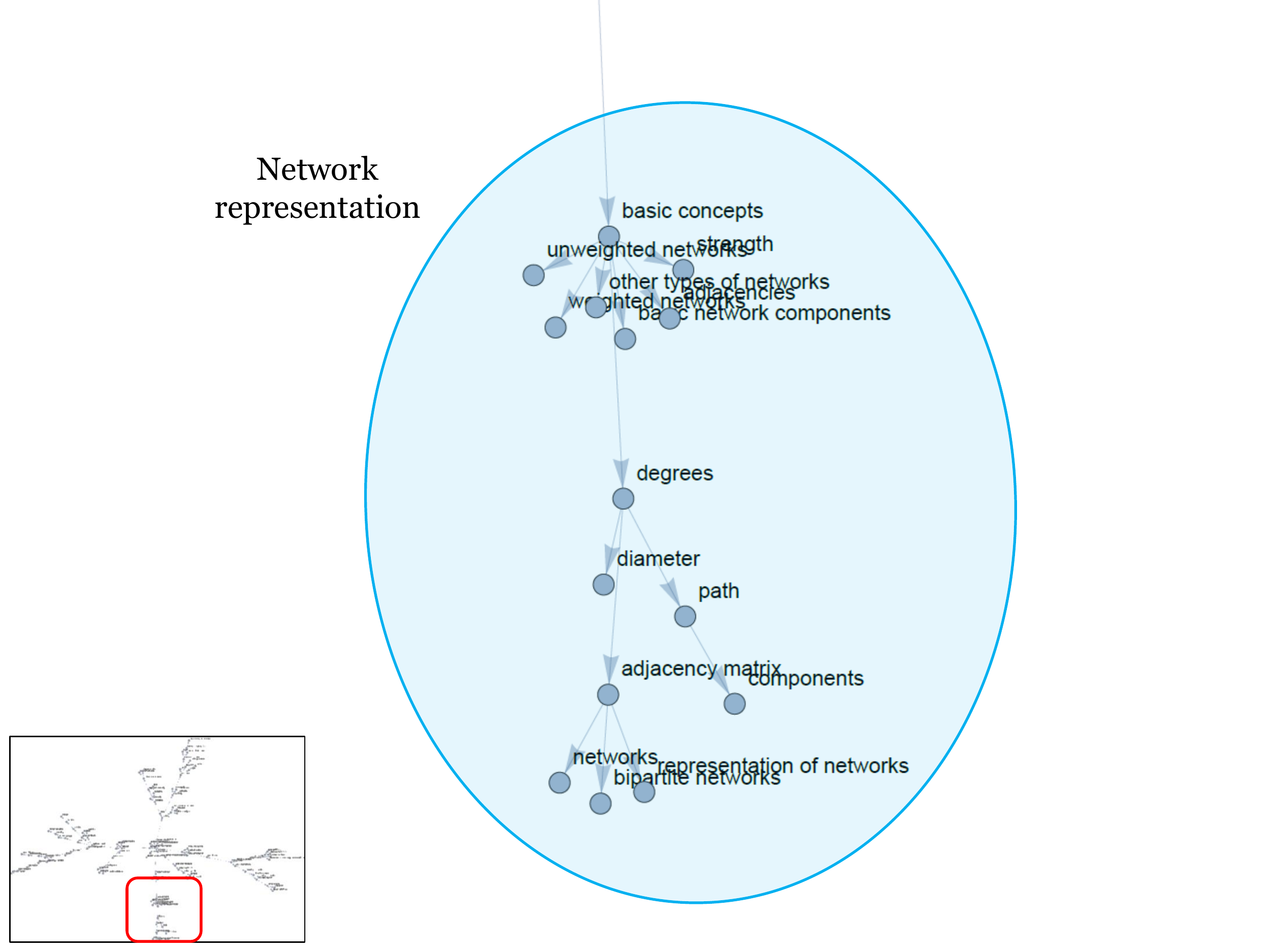}
\caption{Enlarged bottom portion of the spanning tree shown in
  Fig.~\ref{fig:whole-map}. This branch includes various topics about
  {\em network representation.}}
\label{fig:map3}
\end{figure}

Figure \ref{fig:map3} shows the bottom portion of
Fig.~\ref{fig:whole-map} that can be summarized as a single branch
about {\em network representation}, where fundamental concepts and
terminologies about representation of networks are covered, such as
degrees, strengths, adjacencies, unweighted/weighted networks, path,
diameter, and bipartite networks.

\begin{figure}[tbp]
\centering
\includegraphics[width=\columnwidth]{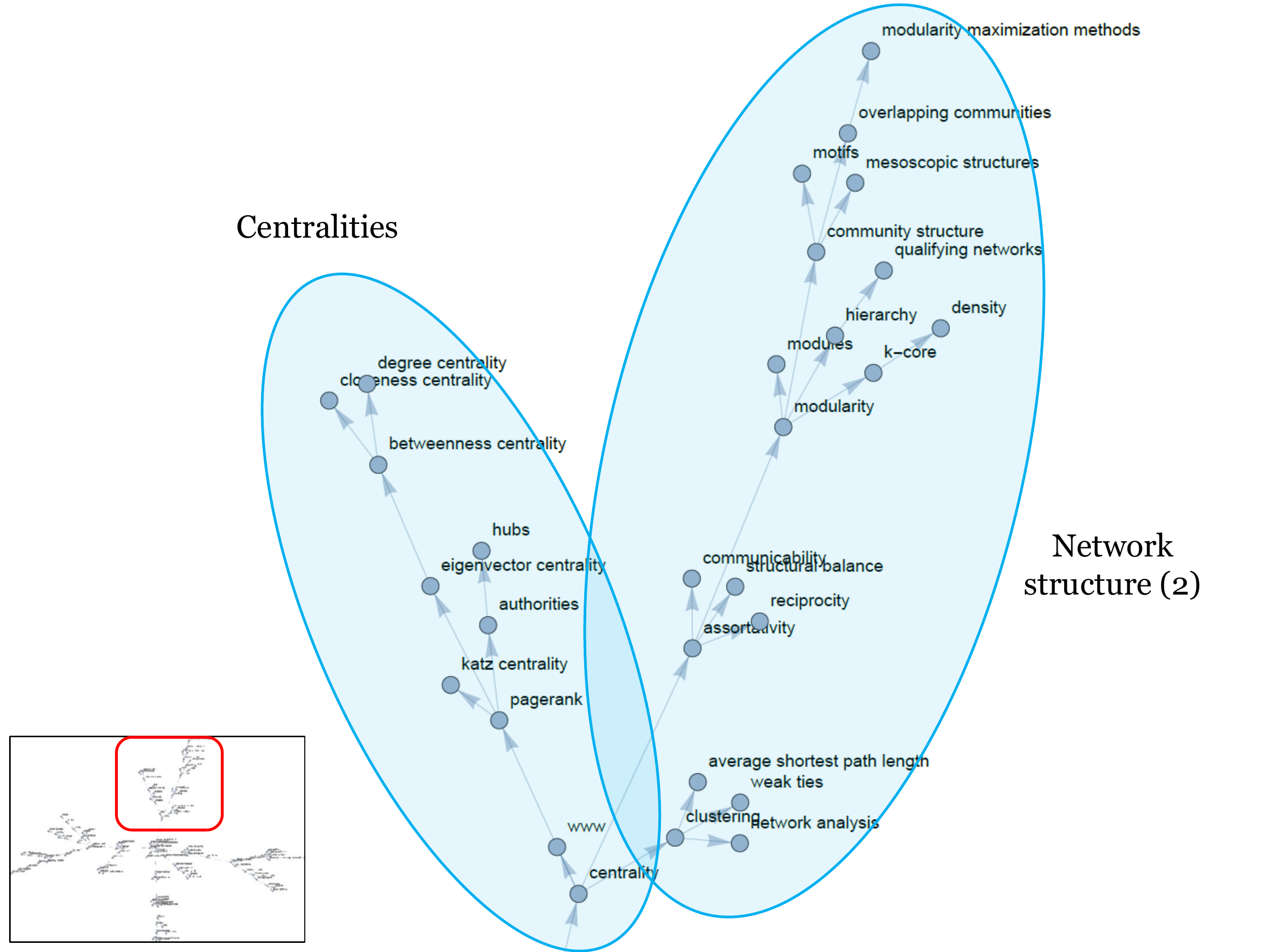}
\caption{Enlarged top portion of the spanning tree shown in
  Fig.~\ref{fig:whole-map}. This portion first creates a major branch
  of {\em centralities}, and then creates another on {\em network
    structure} that covers topics such as assortativity, modularity,
  and community structure.}
\label{fig:map4}
\end{figure}

Figure \ref{fig:map4} shows the top portion of
Fig.~\ref{fig:whole-map}, which includes a branch for {\em
  centralities} and another branch for {\em network
  structure}. Together with the bottom branch shown in
Fig.~\ref{fig:map3}, these three branches cover various topics about
theories and methods of structural analysis of networks.

\begin{figure}[tbp]
\centering
\includegraphics[width=\columnwidth]{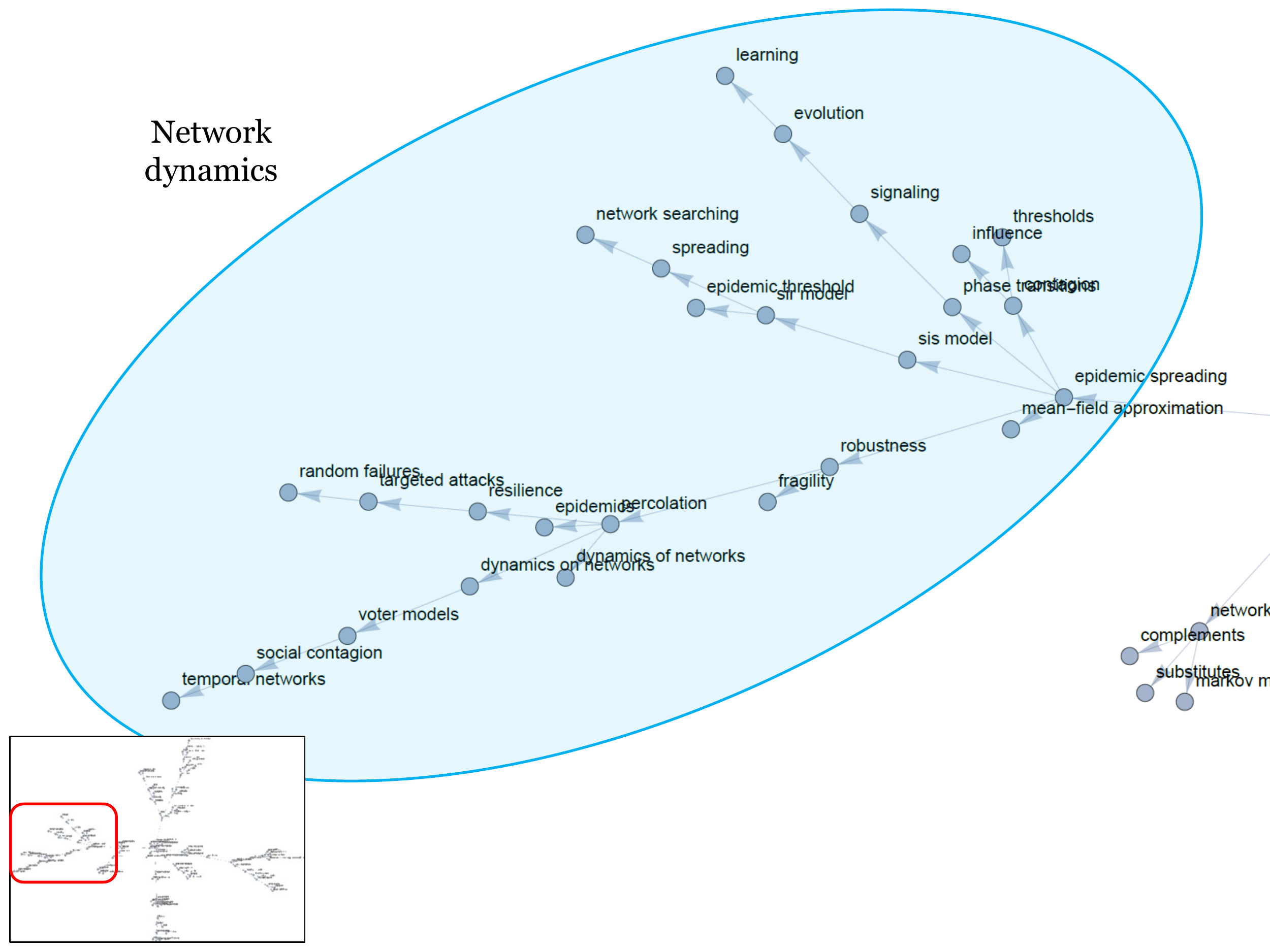}
\caption{Enlarged left portion of the spanning tree shown in
  Fig.~\ref{fig:whole-map}. This portion, coming from {\em diffusion
    \& influence} in Fig.~\ref{fig:map2}, includes a wide variety of
  topics about {\em network dynamics}, such as epidemic spreading,
  phase transitions, robustness, percolation, and dynamics on/of
  networks.}
\label{fig:map5}
\end{figure}

Finally, Figure \ref{fig:map5} shows the left portion of
Fig.~\ref{fig:whole-map}, which can be considered a large branch of
{\em network dynamics}. Extending from {\em diffusion \& influence} in
Fig.~\ref{fig:map2}, this branch covers topics such as epidemic
spreading, phase transitions, robustness, percolation, and dynamics
on/of networks. It is apparent that this area is predominantly
oriented to dynamical systems, stochastic/probabilistic systems, and
statistical physics, where many advanced concepts, theoretical models,
and analytical methods are discussed.

Overall, the examination of the spanning tree illustrated the
following steps as a potential curricular flow of network science
courses:
\begin{enumerate}
\item Start with examples of networks (e.g., social networks), with
  some basics of network structure.
\item Introduce random networks and small-world networks.
\item From there, take any of the following subtopic paths depending
  on the objective and need of the course:
\begin{enumerate}
\item Scale-free networks and network growth
\item Network representation
\item Centralities
\item Other topics on network structure
\item Network dynamics
\end{enumerate}
\end{enumerate}
Needless to say, this presents nothing more than just one example of a
number of possible instruction designs in teaching network
science. Many of the courses included in the dataset of this study
adopted a curricular flow substantially different from the one shown
above (for example, see \cite{porter2017course}). The curricular flow
of a specific course should be carefully custom-designed according to
the objective and scope of the course, the academic level and
background of students, time/resource constraints, and many other
variables.

\section{Conclusions}

In this study, we constructed and analyzed networked maps of topics
covered in 30 existing network science courses. The communities
identified in the topic network revealed seven major topic clusters:
examples of networks, network representation, random networks, network
structure, centralities, network dynamics, and others. These detected
clusters showed a reasonable level of agreement with the essential
concepts identified in the Network Literacy initiative, although the
importance of visualization and computer technology was not well
represented in the current network science courses. This presents a
potential room for instructional redesign; increasing the time and
resource allocated for visualization and computer technology may
improve students' learning of networks.

We also computed the minimum spanning tree of the topic network to
elucidate instructional flows of curricular contents. This analysis
revealed a more fine-grained, directed structure of the topic network,
in which a typical flow of instruction starts with examples of
networks, moves onto random networks and small-world networks, and
then branches off to various subtopics from there. This directed topic
map will be useful for instructors to navigate through various network
science topics and design their own curricula when teaching network
science. We hope that the results presented in this chapter offers the
first step to illustrate the current state of consensus formation
(including variations and disagreements) in the network science
community, on what should be taught about networks and how. They may
also be informative for K--12 education and informal education as
well, when educators and students explore network science topics to
choose relevant teaching/learning materials for their needs.

It should be noted that our results depend on the specific choices we
made about data sources and data cleaning/analysis methods, which were
not fully validated in an objective manner. Conducting a similar
analysis using different sources and methods may thus produce
significantly different maps of curricular contents. Moreover, as the
educational effort of network science has been growing rapidly
\cite{sayama2017netscied}, new courses are continuously created and
offered with new topics, instructional designs, and methodologies each
year. We suggest that the network science community should continue
modeling and analyzing the curricular structure of network science
courses in the coming decades, to develop, assess and adjust effective
teaching strategies and methods for this quickly evolving field of
interdisciplinary research.

\section*{Acknowledgments}

The author thanks Mason Porter for reviewing the earlier version of this
chapter and providing many valuable comments and suggestions, which
have significantly helped improve the content.

\bibliographystyle{spmpsci}
\bibliography{sayama}

\end{document}